\documentclass[preprint,superscriptaddress,nofootinbib]{revtex4}
\usepackage[dvips]{graphicx}
\usepackage[usenames]{color}
\usepackage{setspace}
\usepackage{amsmath}
\usepackage{amssymb}
\usepackage{amsthm}
\usepackage{epsf}
\usepackage{epsfig}
\usepackage{hyperref}

\begin{document}


\preprint{UT-HET-119}

\title{
  Thermal relic abundance of the lightest Kaluza-Klein particle
  in phenomenological universal extra dimension models
}
\author{Yoshiaki~Ishigure}
\email{ishigure@jodo.sci.u-toyama.ac.jp}
\affiliation{
  Department of Physics, University of Toyama, 
  3190 Gofuku, Toyama 930-8555, Japan
}
\author{Mitsuru~Kakizaki}
\email{kakizaki@sci.u-toyama.ac.jp}
\affiliation{
  Department of Physics, University of Toyama, 
  3190 Gofuku, Toyama 930-8555, Japan
}
\author{Akiteru~Santa} 
\email{santa@jodo.sci.u-toyama.ac.jp}
\affiliation{
  Department of Physics, University of Toyama, 
  3190 Gofuku, Toyama 930-8555, Japan
}


\begin{abstract}
  
Universal extra dimension models with Kaluza-Klein parity provide us excellent
candidates for dark matter.
We consider phenomenological universal extra dimension models where
the Kaluza-Klein (KK) mass spectrum is different
from that of the minimal universal extra dimension model,
and compute the thermal relic abundance 
of the first KK mode of 
the photon taking into account the production of
second KK particles.
It is pointed out that its thermal relic abundance depends
significantly on the mass degeneracy between
the KK-photon and other KK particles
because of considerable coannihilation effects.
The cosmologically favored compactification scale is shown
to range from around $1$~TeV to a few TeV even in the cases
where one of the first KK particles is tightly degenerate with the first 
KK photon in mass.

\end{abstract}


\maketitle




The discovery of the Higgs boson with a mass of $125$~GeV 
at the CERN Large Hadron Collider (LHC), 
which was announced on July 4th, 2012~\cite{Aad:2012tfa,Chatrchyan:2012xdj},
is one of the most notable 
breakthroughs in the last decades.
It has been clearly established that elementary particles
in the Standard Model (SM) acquire mass through the Higgs mechanism.
Meanwhile, no clear collider signatures that demand extensions 
of the SM have been found at the LHC.
On the other hand, cosmological and astrophysical observations 
have accumulated evidence of new physics beyond the SM (BSM).
Such BSM phenomena include the existence of dark matter (DM),
the baryon asymmetry of the Universe,
the cosmic inflation as well as neutrino oscillations.
Revealing these phenomenological problems 
by extending the SM is a matter of primary importance.


As for issues concerning DM,
one of the most promising candidates for DM is
a Weakly Interacting Massive Particle (WIMP).
It is remarkable that the existence of a terascale new physics
is inferred from the condition that 
the thermal WIMP relic abundance coincides with the 
DM abundance $\Omega h^2 = 0.1$ determined most notably by
the WMAP~\cite{Hinshaw:2012aka} and Planck~\cite{Ade:2015xua} observations.
This energy scale is now explored by
the ongoing LHC Run-II and will be scrutinized by the
future electron-positron colliders, such as International
Linear Collider (ILC)~\cite{ILC,Asner:2013psa,Moortgat-Picka:2015yla,Fujii:2015jha}, the Compact LInear Collider (CLIC)~\cite{CLIC} 
and the Future Circular Collider of electrons and positrons 
(FCC-ee)~\cite{FCC-ee}, 
as well as DM direct and indirect detection experiments.
Therefore, it is of particular importance to narrow down the energy scale
of new physics models through the precise computation of the WIMP abundance
for such experiments.

In this Letter, as a new paradigm realized at the terascale,
we address models with Universal Extra Dimensions (UEDs)
where all the SM particles freely
propagate in the bulk of extra dimensions~\cite{Appelquist:2000nn}.
For reviews, see, for 
example Refs.~\cite{Hooper:2007qk,Servant:2014lqa,Cornell:2014jza}.
It is intriguing that in UED models equipped with 
Kaluza-Klein (KK) parity,
the Lightest KK Particle (LKP) is stabilized and serves as a 
candidate for WIMP DM~\cite{Cheng:2002ej}. 
One of the simplest UED models is the so-called the minimal UED (mUED) model, 
which assumes the existence of a flat fifth dimension 
compactified on an $S^1/Z_2^{}$ orbifold with a compactification radius $R$
in order to obtain chiral fermions.
In the mUED model, it is also hypothesized that brane localized operators
are absent at the cutoff scale $\Lambda$,
and that mass splitting among KK particles is attributed to 
radiative corrections~\cite{Cheng:2002iz}.
Therefore, physical observables in the mUED model are controlled solely by
the compactification scale $1/R$ and the cutoff scale $\Lambda$.
The LKP in the mUED model is the first KK mode
of the photon $\gamma^{(1)}$, which is the lighter eigenstate of the admixtures 
of the first KK modes of
the $U(1)_Y^{}$ $B$-boson and the neutral $SU(2)_L^{}$ $W^3$-boson.
Since KK particles are almost degenerate in mass at each KK level in the 
mUED model, a large number of coannihilation processes must 
be taken into account in evaluating the relic abundance of $\gamma^{(1)}$.
The complicated computation of the relic abundance of $\gamma^{(1)}$
has been developed in Refs.~\cite{Servant:2002aq,Kakizaki:2005en,Kakizaki:2005uy,Burnell:2005hm,Kong:2005hn,Kakizaki:2006dz,Belanger:2010yx}.
In particular, it has been pointed out that the production of
second KK particles gives the dominant contribution to the effective
annihilation cross section~\cite{Belanger:2010yx}.
The phenomenology of the mUED model
has been extensively investigated from many aspects.
Recent related 
works include Higgs phenomenology~\cite{Belanger:2012mc,Kakuda:2013kba}, 
DM direct detection~\cite{Belanger:2010yx,Hisano:2010yh},
DM indirect detection~\cite{Arina:2014fna}
and collider tests~\cite{Murayama:2011hj,Belyaev:2012ai,Choudhury:2016tff}.
The strongest experimental bound on the mUED parameter space is obtained 
through the LHC Run-II 3.2~$\mathrm{fb}^{-1}$ multijets plus missing transverse
energy searches as $R^{-1}>1100$~GeV insensitive to 
$\Lambda R$~\cite{Choudhury:2016tff}.
On the other hand, it should be noticed that the mUED model is 
constructed based merely on minimality.
Even in the framework of five-dimensional (5D) space-time,
many extended UED models have been proposed including 
non-minimal UED models~\cite{Flacke:2008ne} and split UED 
models~\cite{Park:2009cs}.
In addition, 
the form of renormalization group equations 
and resulting KK mass spectra are easily affected
by introducing new multiplets below the cutoff scale $\Lambda$.
The phenomenological and cosmological consequences of such extended UED models
are quite different from those of the mUED model.

The goal of this paper is to investigate the thermal relic abundance of 
the $\gamma^{(1)}$ LKP in 5D
phenomenological UED (pUED) models where the masses
of KK particles are arbitrary.
Particular attention is paid to the cases where the above-mentioned
second KK particle production is significant.
We show that the compactification scale consistent with the measured
dark matter abundance is considerably dependent on the mass differences
among KK particles.


We first briefly review the framework of 5D UEDs with
the spatial extra dimension compactified on an $S^1/Z_2^{}$ orbifold
whose compactification radius is $R$, and
resulting phenomenological consequences.
The range of the fifth coordinate $y$ is $0 \leq y < \pi R$
because $y$ and $- y$ are identified by orbifolding.
All the SM fields mediate in the bulk of the flat 
extra dimension, resulting in KK tours truncated at the cutoff scale $\Lambda$
in the four-dimensional (4D) viewpoint.
In the mUED, 
the conservation of local 5D Lorentz symmetry and
the absence of brane-localized operators are assumed 
at the cutoff scale $\Lambda$, where
$n$th KK particles have a mass of $n/R$ up to contributions from 
the Higgs vacuum expectation value $v=246$~GeV.
However, note that 
the $S^1$ compactification violates the 5D Lorentz invariance, and that
the $Z_2^{}$ orbifolding does 
momentum conservation along the fifth dimension.
Therefore, radiative corrections generate bulk and
brane-localized operators that give rise to
mass splitting among KK particles at the same KK level.
Although the KK number conservation is violated,
the KK parity defined by $P=(-1)^n$ is still a conserved quantum number.
Consequently, the LKP is stabilized and can become a candidate for DM.
In the mUED, the LKP is the first KK photon $\gamma^{(1)}$,
which is the lighter mass eigenstate of the first $U(1)_Y^{}$ gauge boson, 
$B^{(1)}$, and the neutral component of the first $SU(2)$ 
gauge bosons, $W^{3(1)}$.
In the mass matrix in the $(B^{(1)},W^{3(1)})$ basis,
the diagonal components are substantially larger than the off-diagonal ones,
which are in proportion to $v^2$,
because $1/R \gg v$.
In practice, the weak mixing angle of the first KK gauge bosons 
is small enough to take $\gamma^{(1)} \simeq B^{(1)}$.
It has been shown that 
the effective annihilation cross section of the $\gamma^{(1)}$ LKP
is significantly enhanced by the production processes of second
KK particles and that the mass of the $\gamma^{(1)}$ LKP
consistent with the DM abundance is pushed above 
$1$~TeV in the mUED model~\cite{Belanger:2010yx}.
This is because the second KK production processes
occur near a pole as $m^{(2)} \simeq 2 m^{(1)}$, 
and some of the produced second KK particles decay dominantly
into a pair of SM particles.

\begin{figure}[t]
  \begin{center}
    \includegraphics[clip]{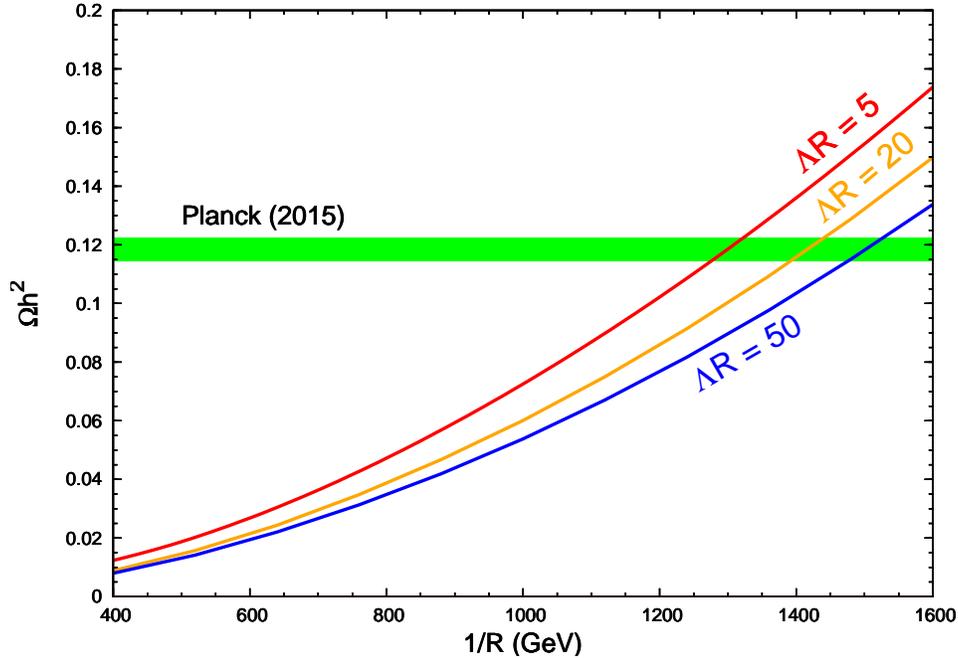}
    \caption{
      The $\gamma^{(1)}$ thermal relic abundance $\Omega h^2$ 
      as a function of the compactification scale $1/R$ in the mUED model.
      The cases with $\Lambda R=5$ (red line), $20$ (orange) and $50$ (blue) 
      are plotted from the top.
      The green band shows the DM abundance at the $2\sigma$ level
      determined by Planck.
}
    \label{fig:mUED}
  \end{center}
\end{figure}

Fig.~\ref{fig:mUED} shows
the thermal relic abundance of the first KK photon $\Omega h^2$ as
a function of the compactification scale $1/R$ in the mUED model.
The cases with $\Lambda R=5$ (red line), $20$ (orange) and $50$ (blue) are 
plotted from the top.
The green band shows the DM abundance at the $2\sigma$ level
determined by Planck~\cite{Ade:2015xua}.
Here, we include all the coannihilation processes taking 
the production of second KK particles into account.
In computing the KK mass spectrum and 
the thermal relic abundance of $\gamma^{(1)}$, 
we use the same model files as Ref.~\cite{Belanger:2010yx} employs.
Namely, these model files are generated by {\tt LanHEP}~\cite{Semenov:2014rea} 
and implemented into {\tt CalcHEP}~\cite{Belyaev:2012qa} 
and {\tt micrOMEGAs4.3}~\cite{Barducci:2016pcb}.
For the details of the computations, see Ref.~\cite{Belanger:2010yx}.
The mass of the Higgs boson is set at $m_h^{}=125$~GeV.
From this figure, the cosmologically favored 
compactification scale, which is roughly the mass of the fist KK photon 
$\gamma^{(1)}$, is shown to range typically 
from around $1300$~GeV to $1500$~GeV in the mUED model.

We analyze the thermal relic density of the $\gamma^{(1)}$ LKP in pUED models.
We emphasize that the mUED model presented above is just a benchmark model
of the framework of 5D UEDs on an $S^1/Z_2^{}$.
Namely, the mUED is the UED counterpart of
the constrained Minimal Supersymmetric Standard Model (cMSSM)
in that the number of free parameters is minimized.
In general, UED models offer us a variety of phenomenological
consequences.
Baring this situation in mind, we consider pUED models that allow
arbitrary KK masses retaining the feature that the mass of the $n$th KK
particle is $n$ times that of the first KK particle in the same 5D field.
To this end, 
we introduce new field strength parameters $Z_A^{}$, 
$Z_{\psi}^{}$, and $Z_{\Phi}^{}$ 
in the kinetic terms 
of the gauge bosons, fermions and Higgs boson in the 5D Lagrangian,
\begin{eqnarray}
  \mathcal{L}^{(5D)}_{A} &=& -\frac{1}{4}F^a_{\mu\nu}F^{a\mu\nu}
                            -\frac{1}{2} Z_A^{} F^a_{\mu5}F^{a\mu5},\\
  \mathcal{L}^{(5D)}_{\psi} &=& \bar{\psi}i \gamma^{\mu}D_{\mu}^{} 
                                - Z_{\psi}\bar{\psi}i \gamma^{5}D_{5}^{},\\
  \mathcal{L}^{(5D)}_{\Phi} &=& (D^{\mu} \Phi)^{\dagger}(D_{\mu}^{}\Phi) 
                                - Z_{\Phi}(D_5^{} \Phi)^{\dagger}(D_{5}^{}\Phi)
                                - \mu^2 \Phi^{\dagger}\Phi,
\end{eqnarray}
respectively.
Before the electroweak symmetry breaking, 
the (squared) masses of $n$th KK particles of the gauge bosons, 
fermions and Higgs boson
are given by $m_{A^{(n)}}^2 = Z_A^{} n^2/R^2$, 
$m_{\psi^{(n)}}^{} =Z_{\psi} n/R$, 
and $m_{\Phi^{(n)}}^2 = Z_{\Phi} n^2/R^2 + \mu^2$, respectively.
If we fix the values of these field strength parameters to those
determined solely by 
the radiative corrections from the SM particles and their KK modes,
we recover the mUED results.
Instead of taking the mUED values,
we regard the field strength parameters as arbitrary
and investigate to what extent the thermal relic density of
the first KK photon is changed.
It has been known that in UED models the inclusion of
coannihilation processes can significantly affect 
the resultant relic density~\cite{Servant:2002aq,Kakizaki:2005en,Kakizaki:2005uy,Burnell:2005hm,Kong:2005hn,Kakizaki:2006dz}.
In discussing the coannihilation effects, it is convenient to introduce
the following mass degeneracy parameter:
\begin{equation}
  \Delta_X^{} = \frac{m_{X^{(1)}}^{} - m_{\gamma^{(1)}}^{}}{m_{\gamma^{(1)}}^{}}.
\end{equation}
We focus on the cases where only the mass of one of the first KK particles 
is taken arbitrary and set the others to the mUED values.
Since the radiative corrections to the $B^{(1)}$-boson are negligible
in the mUED, we obtain $Z_{B}^{} \simeq 1$.
Therefore,
the mass degeneracy parameter for $X$ is roughly rewritten as 
$\Delta_X^{} \simeq Z_X^{} -1$.
As in the mUED model, we set the bulk Higgs mass parameter at $\mu=0$.

\begin{figure}[t]
  \begin{center}
    \includegraphics[clip]{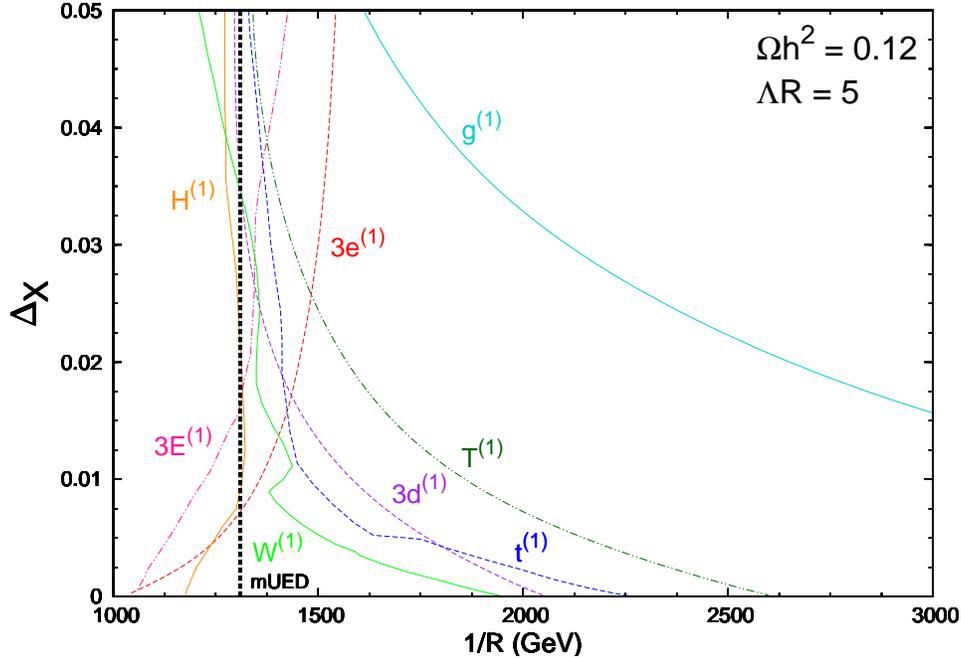}
    \caption{
The contours of the thermal relic abundance
of the first KK photon satisfying
$\Omega h^2=0.12$ in the $(1/R,\Delta_X^{})$ plane.
We show the degenerate cases where the first KK particle of $X$ is
the first KK gluon $g^{(1)}$ (cyan line), 
the first KK $W$-bosons $W^{(1)}$ (light green), 
the first KK left-handed top quark $T^{(1)}$ (dark green), 
the first KK right-handed top quark $t^{(1)}$ (blue), 
the three generations of the first 
KK right-handed down-type quarks $3d^{(1)}$ (purple), 
the three generations of the first 
KK left-handed leptons $3E^{(1)}$ (pink), 
the three generations of the first 
KK right-handed leptons $3e^{(1)}$ (red),
and the first KK Higgs bosons $H^{(1)}$ (orange).
The other mass parameters than $X$ 
are set to the mUED values with $\Lambda R=5$.
For comparison, the mUED prediction (black) is also shown.
    }
    \label{fig:pUED}
  \end{center}
\end{figure}

Fig.~\ref{fig:pUED} shows the contours of the thermal relic abundance
of the first KK photon satisfying 
$\Omega h^2=0.12$ in the $(1/R,\Delta_X^{})$ plane.
We investigate the degenerate cases where the first KK particle of $X$ is
the first KK gluon $g^{(1)}$ (cyan line), 
the first KK $W$-bosons $W^{(1)}$ (light green), 
the first KK left-handed top quark $T^{(1)}$ (dark green), 
the first KK right-handed top quark $t^{(1)}$ (blue), 
the three generations of the first 
KK right-handed down-type quarks $3d^{(1)}$ (purple), 
the three generations of the first 
KK left-handed leptons $3E^{(1)}$ (pink), 
the three generations of the first 
KK right-handed leptons $3e^{(1)}$ (red),
and the first KK Higgs bosons $H^{(1)}$ (orange).
The other mass parameters than $X$ 
are set to the mUED values with $\Lambda R=5$.
As in the mUED case,
we take into account all the contributions 
from the coannihilation modes with the other 1st KK particles
including the production of 2nd KK particles in the final state.
The forms of the KK-number violating vertices, 
which make the second KK particles decay, 
are set to those in the mUED model for simplicity.
The nontrivial behavior for the $Z^{(1)}$ line stems from whether or not
the mass of the second KK charged $W$-bosons, $W^{(2)\pm}$,
is larger than the sum of those of $Z^{(1)}$ and
the first KK charged Higgs bosons, $a^{(1)\pm}$.
For comparison, the mUED prediction (black) is also shown.
From this figure, it clear that in general the predicted compactification
scale varies significantly depending on the mass degeneracy parameters.
Even in these examples, the allowed compactification scale ranges from 
around $1$~TeV to a few TeV.

In conclusion, 
we have evaluated the thermal relic abundance 
of the first KK photon $\gamma^{(1)}$ including the production of
second KK particles in pUED models
where KK mass shifts are taken arbitrary.
We have shown that the thermal relic abundance of 
the $\gamma^{(1)}$ LKP depends
crucially on the mass degeneracy between $\gamma^{(1)}$ and other KK particles
as coannihilation effects are significant.
The range of the compactification scale consistent with the observed 
DM abundance
has been shown to be from around $1$~TeV to a few TeV if
the mass of one of the first KK particles is tightly degenerate with 
that of the $\gamma^{(1)}$ LKP.

\begin{acknowledgments}
This work was supported, in part, by 
Grant-in-Aid for Scientific Research on Innovative Areas,
the Ministry of Education, Culture, Sports, Science and Technology,
No.\ 16H01093 (M.K.),
and by 
the Sasakawa Scientific Research Grant from The Japan Science Society (A.S.).
\end{acknowledgments}

\end{document}